\documentclass[twoside,12pt]{article}
\usepackage{epsfig}

\def\Journal#1#2#3#4{{#1} {#2} (#4) #3 }

\def\PLB{{\em Phys. Lett.} B}

\def\PRL{\em Phys. Rev. Lett.}

\def\PRD{{\em Phys. Rev.} D}

\newcommand{\be}{\begin{equation}}
\newcommand{\ee}{\end{equation}}
\newcommand{\bea}{\begin{eqnarray}}
\newcommand{\eea}{\end{eqnarray}}

\begin{document}

\title{\vspace{1cm}Low energy neutrino astronomy with the large liquid
scintillation detector LENA}

\author{T. Marrod\'an Undagoitia$^{1}$\footnote{Corresponding
authors. Fax: +49 89 289 12680, Email: tmarroda@ph.tum.de and
franz.vfeilitzsch@ph.tum.de}, F. von Feilitzsch$^{1\ast}$,
M. G\"oger-Neff$^1$,\\ K. A.  Hochmuth$^2$, L. Oberauer$^1$,
W. Potzel$^1$, M. Wurm$^1$\\ \\$^1$Physik-Department E15, Technische
Universit\"at M\"unchen,\\ James-Franck-Str., 85748 Garching,
Germany\\ $^2$Max-Planck-Institut f\"ur Physik
(Werner-Heisenberg-Institut)\\ F\"ohringer Ring 6, 80805 M\"unchen,
Germany}
\date{Progress in Particle and Nuclear Physics 57 (2006) 283}
\maketitle

\begin{abstract} 
The detection of low energy neutrinos in a large scintillation
detector may provide further important information on astrophysical
processes as supernova physics, solar physics and elementary particle
physics as well as geophysics. In this contribution, a new project for
{\bfseries L}ow {\bfseries E}nergy {\bfseries N}eutrino {\bfseries
A}stronomy (LENA) consisting of a 50~kt scintillation detector is
presented.
\end{abstract}

\pagebreak

\section{Introduction}

Low energy neutrino research was particularly successful in the recent
years in neutrino astronomy and neutrino elementary particle
physics. This lead to the first observation of a supernova explosion
via neutrino detection~\cite{SN1}\cite{SN2}, the measurement of the
solar neutrino spectrum~\cite{Bachcall05} and the discovery of
neutrino oscillations~\cite{GNO}\cite{SNODis}\cite{KamLANDDis}. In
this paper, we investigate the potential for new discoveries if low
energy neutrinos can be detected in a large scintillation detector
with a total mass of 50~kt using the low background technology
developed for the BOREXINO detector~\cite{BorexTech}. These
investigations focus on the following topics:
\begin{enumerate}
\item Solar neutrino spectroscopy,

\item Neutrino detection from a nearby supernova explosion (in the center
of the milky way),

\item Detection of relic neutrinos from previous supernova explosions,

\item Detection of neutrinos emitted from the Earth,

\item Search for proton decay.\\
\end{enumerate}

\section{The LENA detector design}

The {\bfseries L}ow {\bfseries E}nergy {\bfseries N}eutrino {\bfseries
A}stronomy (LENA) detector~\cite{Ob05}\cite{TMarPRD05}, is assumed to
be constructed as a double-walled cylinder with a diameter of 30 m and
a length of approximately 100 m. As indicated in figure 1, the inner
volume is instrumented at the walls with photomultiplier tubes
providing a surface coverage of 30\% photo-sensitive cathode area and
filled with 50~kt of liquid scintillator consisting of PXE
(Phenyl-o-xylylethane) as a solvent and $\sim$~2~g/l of pTp and 20~mg/l
bisMSB serving as a fluor and wavelength shifter. This scintillator
was developed as an option for the BOREXINO
detector~\cite{BorexTech}. The scintillator provides a density of
0.98~g/cm$^3$. With a light transmission length of 12~m a
photoelectron efficiency of $\sim 120$~pe/MeV is expected. This is a
quite realistic value considering the value of $\lambda=12$~m measured
at the Counting Test Facility (CTF) in Gran Sasso~\cite{CTF}. \\

In the outer section of the cylinder, 2~m of water provide shielding
against external radioactivity, as well as a water Cerenkov detector
serving as a muon veto. This design should make it possible to reach a
sufficiently low internal gamma-induced background to avoid too high a
signal rate at the photomultiplier tubes provided the photomultipliers
are sufficiently low in internal radioactivity.\\

The detector is designed for a detection threshold of 250 keV,
yielding 30~photoelectrons at threshold, and should be constructed at
a deep underground site of more than 4000 m water equivalent to
provide a sufficient muon background reduction. At the same time, it
should be far enough from nuclear power plants to limit the electron
antineutrino background. In Europe, there are two favourable sites
satisfying these requirements. One is in the south of Greece, at the
Nestor deep underwater laboratory where the detector might be placed
at a depth of 4000-5000 m water in the Mediterranean sea. A second
suitable place might be at the deep underground laboratory in
Pyh\"asalmi (CUPP: Center of Underground Physics in Pyh\"asalmi) in a
mine at a depth of 1400 m rock. Both laboratories are situated far
from nuclear power plants. For the Pyh\"asalmi
laboratory~\cite{Pyhasalmi}, the estimated electron antineutrino
background flux from nuclear power plants of about
$6\cdot10^9$~m$^{-2}$~s$^{-1}$ is sufficiently low even if a nuclear
power reactor plant would be constructed in Finland in the future.

\begin{figure}
  \begin{center}\includegraphics[
      width=0.40\columnwidth,
      keepaspectratio,
      angle=0]{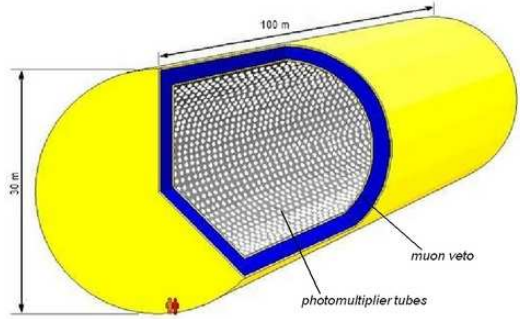}
    \begin{minipage}[t]{16 cm}
      \caption{Artistic illustration of the LENA detector. The inner part of
	about 13~m radius will contain approximately 50~kt of liquid
	scintillator while the outer part ($13-15$~m radius) will be filled
	with water acting as muon-veto.\label{fig1}}
    \end{minipage}
  \end{center}
\end{figure}

\section{Detection of solar neutrinos}

The detector LENA would be able to detect solar $^{7}$Be neutrinos via
neutrino electron scattering with a rate of $\sim
5400$~events/day. Depending on the signal-to-background ratio, this
would provide a sensitivity for time variations in the $^{7}$Be
neutrino flux of $\sim 0.5$\% during one month of measuring time. Such
a sensitivity may give information at a unique level on
helioseismology (pressure or temperature fluctuations) and on a
possible magnetic moment interaction with a timely varying solar
magnetic field.\\

The additional detection of pep neutrinos (see figure 2) with a rate
of $\sim210$~events/day could provide information on neutrino
oscillation which is expected to show a transition from matter-induced
oscillation to vacuum oscillation in the energy range between
$1-2$~MeV. The ratio of the neutino flux from the reactions of pep
fusion and pp fusion is theoretically determined with an accuracy
$\leq 1$\%. Thus the measurement of the pep neutrino flux is
effectively as well determining the pp solar neutrino flux.\\

The neutrino flux from the CNO cycle is theoretically predicted only
with the lowest accuracy (30\%) of all solar neutrino
fluxes. Therefore, LENA would provide a new opportunity for a detailed
study of solar physics on a level which will exceed the information
provided by BOREXINO from the statistics point of view by a factor of
more than 100 provided the extremely high radiopurity of the
scintillator expected for BOREXINO will as well be reached in LENA.
\\

With a liquid scintillator, $^{13}$C atoms naturally contained in it
can be used as target for $^{8}$B neutrinos~\cite{13C} as the energy
threshold is 2.2~MeV. Around 360~events of this type per year can be
estimated for LENA. A deformation due to the MSW-effect should be
observable in the low-energy regime of the $^8$B neutrino spectrum
after a couple of years of measurements.

\begin{figure}
  \begin{center}\includegraphics[
      width=0.38\columnwidth,
      keepaspectratio,
      angle=-90]{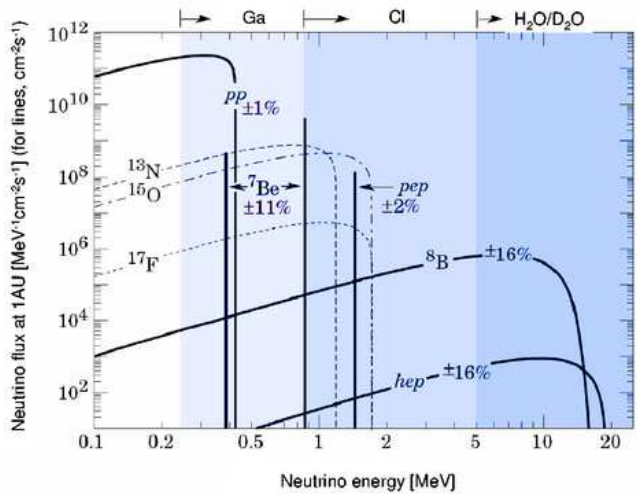}
    \begin{minipage}[t]{16 cm}
      \caption{The solar neutrino energy spectrum on the Earth as predicted
	by the current standard model~\cite{Bachcall05}. Contributions from
	the pp-chain branchen are plotted as solid lines, the CNO
	contributions as dotted lines. The top part of the diagram illustrates
	from left to right the energy threshold for the gallium (GALLEX, GNO,
	SAGE), chlorine (Homestake), as well as water (Superkamiokande) and
	heavy water (SNO) experiments.\label{fig2}}
    \end{minipage}
  \end{center}
\end{figure}

\section{Supernova neutrinos}

A 8~M$_\odot$ supernova exploding in the centre of the milky way
typically will induce a signal rate of $\sim20~000$~events. This would
include neutrinos and antineutrinos of all flavours. The rates of the
different reaction channels are listed in table 1 and have been
obtained by scaling the predicted rates
from~\cite{SNBorex}\cite{SNelasP} to LENA. A discrimination between
electron neutrinos and electron antineutrinos would be possible by the
interaction of antineutrinos via inverse beta decay and neutrino
interaction at $^{12}$C.  The first three are charged current
reactions that will deliver information on $\nu_e$ and
$\overline{\nu_{e}}$ fluxes and spectra. The last three are neutral
current reactions, sensitive to all neutrino flavours. These reactions
give information on the total flux. \\

\begin{table}
\begin{center}
\begin{minipage}[t]{16 cm}
\caption{Possible neutrino reactions in a liquid scintillation type
detector and expected event rates for a 8~$M_{\odot}$ supernova in the
center of our Galaxy ($\sim 10$~kpc distance). For the first five
reactions, see reference~\cite{SNBorex}, for the last reaction, see
reference~\cite{SNelasP}. In both cases, the event rates calculated
for other detectors were scaled to those in LENA taking into account
the number of target atoms in the fiducial volume.}
\label{tab:obe}
\end{minipage}
\begin{tabular}{r|l}
&\\[-2mm]
\hline
&\\[-2mm]
Reactions & Event rate \\
&\\[-2mm]
\hline
&\\[-2mm]
$\overline{{\nu}}_{e}+p\to n+e^{+}$ &$\sim 8700$ \\[2mm]
$\overline{{\nu}}_{e}+{}^{12}\textrm{C}\to {}^{12}\textrm{B}+e^{+}+$& $\sim494$\\[2mm]
${\nu}_{e}+{}^{12}\textrm{C}\to e^{-}+{}^{12}\textrm{N}$& $\sim85$\\[2mm]
${\nu}_{x}+{}^{12}\textrm{C}\to {}^{12}\textrm{C}^{\ast}+{\nu}_{x}$& $\sim2925$\\[2mm]
${\nu}_{x}+e^{-}\to {\nu}_{x}+e^{-}$& $\sim610$\\[2mm]
${\nu}_{x}+p\to {\nu}_{x}+p$&$\sim7370$\\[2mm]
&\\[-2mm]
\hline
\end{tabular}
\end{center}
\end{table}

 Due to the high statistics of the neutrino detection a time-resolved
neutrino-flux rate for different neutrino interactions could be
measured which would give new information on the dynamics of supernova
explosions and would be complementary to the information accessible in
large water-Cerenkov detectors or in a large liquid-Argon detector. A
calculated neutrino time spectrum for a typical type II supernova
explosion is depicted in fig 3.~\cite{Supernova1}\cite{Supernova2}\\

\begin{figure}
  \begin{center}\includegraphics[
      width=0.5\columnwidth,
      keepaspectratio,
      angle=0]{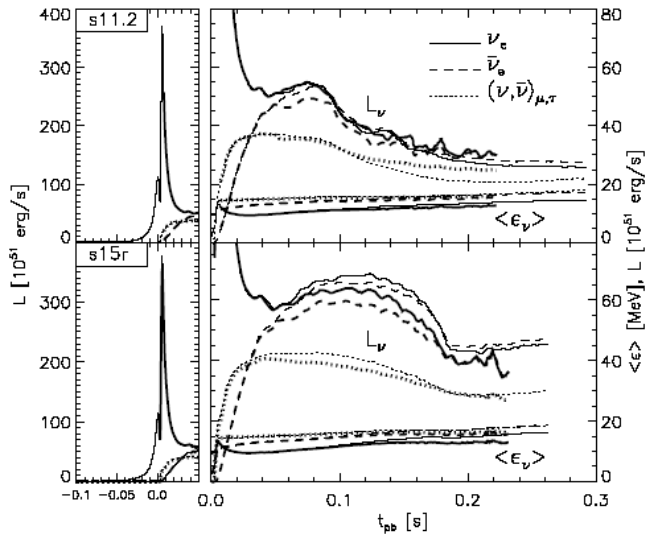}
    \begin{minipage}[t]{16 cm}
      \caption{Luminosities and mean energies for $\nu_{e}$,
	$\overline{\nu}_{e}$ and $\nu_{\mu, \tau}$, $\overline{\nu}_{\mu,
	  \tau}$ vs. time for two different models. The left panels (left
	scale) show the prompt $\nu_{e}$ burst, the right panels enlarge the
	post-bounce time region~\cite{Supernova2}.\label{fig3}}
    \end{minipage}
  \end{center}
\end{figure}

\section{Supernova relic neutrinos}

The detection of relic neutrinos from supernova explosions
(supernova relic neutrinos SRN) in the early universe provide
information on supernova explosion mechanism, the early star-formation
rate and thus, important cosmological data. The current best limit on
SRN-flux comes from the SuperKamiokande experiment, giving an upper
limit of 1.2~cm$^{-2}$s$^{-1}$ for $\overline{\nu}_e$ with an energy
threshold of 19.3~MeV~\cite{SRN-SK}. With LENA, the sensitivity would
be increased, as the inverse beta decay can be used as detection
reaction. This provides a good background rejection and the threshold
energy is expected to be $\sim 10$~MeV.\\

The detection of supernove relic neutrinos is particularly
challenging as it is strongly dependent on the low background
conditions of the detector with respect to electron antineutrinos from
nuclear power reactors. To our knowledge both locations envisioned for
LENA are suitable in this respect. Following the present models of
cosmology, the expected detection rate for these electron
antineutrinos is in the order of several events per year. Here it is
essential to measure the electron antineutrino spectrum down to as low
energies as possible.

\section{Terrestrial neutrinos}

Unique information on the interior of the Earth can be obtained by the
detection of electron antineutrinos from the Earth. Present technology
permits drilling of holes down to a depth of approximately 10 km,
compared to the Earth radius of $\sim6300$~km. It is known that the
emitted heat from the Earth exceeds the energy flow from the Sun by
around $\sim40$~TW. This excess is generally attributed to heat
emitted by natural radioactivity of material inside the
Earth~\cite{Geophysics}. However, estimations of the Earth's interior
composition accounts only for half of this radioactivity. Therefore a
measurement of the electron antineutrino flux could provide valuable
information to solve this puzzle. \\

These so-called geoneutrinos were recently measured for the first time
by the liquid scintillator detector KamLAND~\cite{NatureKam}. In
figure 4, the expected spectra for $^{238}$U, $^{232}$Th and $^{40}$K
decay chain neutrinos are plotted. Neutrinos from $^{40}$K being below
the energy threshold of the inverse beta decay reaction can not be
detected. However, with a good energy-resolution (high statistics),
neutrinos from $^{238}$U and $^{232}$Th decay chain can be
discriminated.\\

In addition, if directional information on the antineutrino flux can
be obtained, a spatial distribution of the natural radioactivity in
the interior of the Earth could be derived. Scaling the experimental
result from KamLAND to LENA detector yields an event rate in the range
between $4\cdot10^2$ and $4\cdot10^3$~y$^{-1}$ for the location in
Pyh\"asalmi (continental crust). In the first calculations, we
overestimated LENA's directional sensitivity based on the
n-displacement after inverse beta decay of the proton due to the
neutrino momentum. These calculations are currently being revised.\\

A natural nuclear reactor has been suggested to reside in the centre
of the Earth thus providing an explanation for the missing energy
source in the Earth~\cite{Georeactor}. The existence of such a nuclear
reactor (for example of $\sim10$~TW) could be tested with LENA.

\begin{figure}
  \begin{center}\includegraphics[
      width=0.40\columnwidth,
      keepaspectratio,
      angle=0]{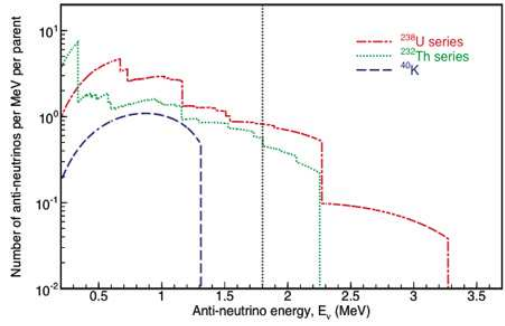}
    \begin{minipage}[t]{16 cm}
      \caption{Expected electron antineutrino energy distributions for the
$^{238}$U, $^{232}$Th and $^{40}$K decay chains~\cite{NatureKam}. In a
liquid-scintillator detector, inverse beta decay in protons can be
used to measure these antineutrinos above the reaction threshold of
1.8~MeV. This threshold is represented with the vertical dotted
line.\label{fig4}}
    \end{minipage}
  \end{center}
\end{figure}

\section{Proton decay}

A large scintillation detector provides unique sensitivity to the
proton decay channel $p\to K^{+}\overline{{\nu}}$. This is due to the
high energy resolution obtained in a scintillator which attains
an approximately 50 times larger light output than water-Cerenkov
detectors at energies below 1~GeV. This decay mode is favoured in many
Supersymmetry theories and may be expected with a lifetime $\tau$
below $10^{35}~\textrm{y}$. The experimental limits on this decay mode
(Super-Kamiokande with
$\tau>2.3\cdot10^{33}~\textrm{y}$~\cite{Super05}) are essentially
restrained by the capability of background suppression or signal
identification in the detector. Based on Monte Carlo calculations for
this decay mode, a lower limit of $\tau>4\cdot10^{34}~\textrm{y}$ at
$90\%$ C.L. has been derived~\cite{TMarPRD05}. More detailed
information on this result is given in a contribution to this
conference by T. Marrod\'an Undagoitia.\\

\section{Summary}
A large scintillation detector with a total detector mass of 50~kt may
provide a unique tool for neutrino astronomy, geophysics and
elementary particle physics. A more detailed design study will be
performed in the near future. The use of a scintillator as a neutrino
target and detector medium has a variety of advantages due to the very
high light output. \\

This detector would be largely complementary to a megaton water
Cerenkov detector being discussed for different locations in Europe,
USA and Japan. Due to a competence and expertise present in Europe,
especially for scintillation detectors with extremely low background
at low energies, a LENA type detector may be of particular interest to
be investigated in Europe.

\section*{Acknowledgments}
We gratefully acknowledge most valuable discussions with E. Lisi. This
work has been supported by funds of the Maier-Leibnitz-Laboratorium
(Garching), the Virtual Institute for Dark Matter and Neutrino Physics
(VIDMAN, HGF) and by the Deutsche Forschungsgemeinschaft DFG
(Sonderforschungsbereich 375).

\end{document}